# Discovering Dynamic Integrity Rules with a Rules-Based Tool for Data Quality Analyzing

Thanh Thoa Pham Thi, Markus Helfert

***Abstract***: *Rules based approaches for data quality solutions often use business rules or integrity rules for data monitoring purpose. Integrity rules are constraints on data derived from business rules into a formal form in order to allow computerization. One of challenges of these approaches is rules discovering, which is usually manually made by business experts or system analysts based on experiences. In this paper, we present our rule-based approach for data quality analyzing, in which we discuss a comprehensive method for discovering dynamic integrity rules.*

## INTRODUCTION

Data quality (DQ) is an increasing concern for most businesses. High quality data helps the organisations to save costs, to make better decisions and to improve customer service. Nowadays, organisations have been increasingly aware of the importance of DQ in their business. In a recent report, Gartner forecasts that "the DQ market will be worth $677 million by 2011, representing a compound annual growth rate of 17.6%".

Typical problems concerning DQ experienced by data-users are inconsistent, incomplete, inaccurate or untimely data. Let us illustrate a typical scenario, concerning DQ problems in e-commerce applications.

Customers choose products and place orders online followed by the product delivery.. The customers have the right to return their products within 15 days after the delivery for change or cancellation purposes, otherwise the order is closed.  In this example, *inconsistent* data can occur when the unit price of a product and the price in the order are not the same, *inaccurate* data may occur when a full price is applied to products on sale, *incomplete* data can happen when some important information is missed such as missing street number and address information in the client address. An example for *untimely* data can be when products are out of stock are displayed for selling.

Addressing DQ problems, several commercial tools and consulting solutions exist for analyzing and improving DQ such as Informatica, Clavis and Trillium software. Key functionalities of these tools are data cleansing and data verification with reference data such as product name, address, telephone format, etc.

Over the last years, business rules (BRs) based approaches for DQ solutions have become promising. For instance SearchDatamanagement web-magazine states that: "the veteran DQ tool vendors are being challenged by entrants that have an international focus and propensity toward designing and deploying domain-agnostic DQ services (…), based on a centrally managed set of BRs" [7]. A BR is "a statement that describes some structural aspect of a business, or defines some relationship between entities in a business, or controls or influences the behaviour of the business" [3]. In this sense, a BR is a *structural assertion* which describes some aspects of enterprise, or an *action assertion* that limits or controls the actions of the enterprise, or a *derivation* which is statement of knowledge derived from other knowledge in the enterprise [3]. Rules based approaches for DQ solutions often use integrity rules for data monitoring purpose.

*Integrity rules* (IRs) are constraints on data that enforce data to be meaningful, consistent, correct and valid according to BRs and they represent BRs in a way that could be stored in rules repository for computerized purpose, because BRs are usually described in natural language. IRs are usually described with predicate logic in If-Then structure, decision tree or decision table, etc.

In generally, BRs can be translated into two kinds of IRs: static and dynamic rules. [2] has defined static rules as rules that can be applied on the data value at any time in the data life-cycle. Dynamic rules concern the data value changes (change data status), which are results of business process transactions. This kind of rules is validated at the moment of data changes. For instance, a static rule is "the order date is earlier or equal to the delivery date of this order", and a dynamic rule states for example "an order is closed (change to state close) after 15 days of delivery and do not have corresponding returned products".

Indeed commercial tool providers, such as Informatica and Clavis follow a rule based approach to DQ. Usually BRs and IRs are analyzed manually by business experts and/or system analysts. Although this manual approach is suitable and straightforward for static rules, for dynamic rules it can be complex and challenging.

In order to address this problem, research has contributed approaches such as mining BRs from event log files in Business Process Management Systems [1]. These approaches concern basically authorization rules and rules on actions (in which condition an action is made). However this approach is out of scope for a DQ tool.

In this paper we present a rule-based approach to analyse DQ applies IRs, in which we propose a comprehensive method to discover dynamic IRs based on object life cycles (OLC). The OLC is described with our meta-modelling concept- the Node-Star structure which is the dynamic part of the IASDO model [4], [5]. By applying the IASDO model we have two main advantages: First it is a formal method which can be easily implemented in our tool, and second it has expressive power which allows to describe more information than other methods [5], [6]. In the following we present our rules-based approach and its framework and illustrate it within the e-commerce example. Particularly we explore the issue of discovering dynamic IRs. Finally we conclude our work and discuss some future work.

**FRAMEWORK FOR RULES-BASED DATA QUALITY ANALYZING**

Our framework has four main components (Figure 1).

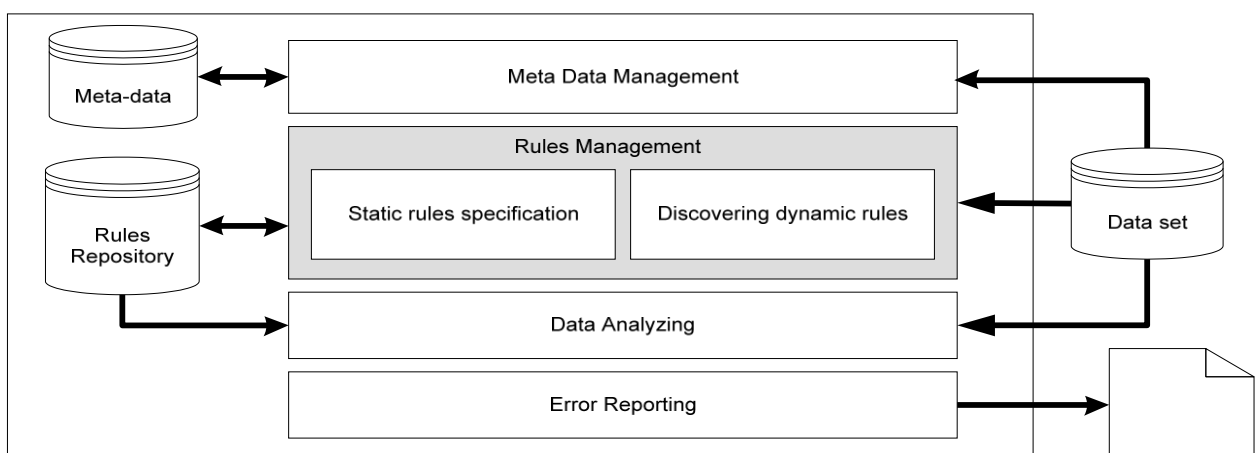

Figure 1. Framework of Rules-based DQ analyzing approach

As external component we illustrate a data set or database as main input. The *Meta-Data Management* component is responsible for capturing the DB Schema and the object life cycle models which are specified by business experts or system analysts The *Rules Management* component manages IRs during the life-cycle and is responsible for generating dynamic rules. The static rules specification can be manually done; meanwhile dynamic rules specification is semi-automatic. The *Data Analyzing* component checks the validity of data according to the rules stored in the rules repository. Finally the *Error reporting* component exports data errors and makes necessary statistics. In this paper, we focus on discovering dynamic IRs. We use e-commerce example to illustrate our approach.

**DISCOVERING DYNAMIC INTEGRITY RULES**

Our method for discovering dynamic IRs is based on the logic of state change defined in object life cycles (OLC), and the logic of state definition in DB schema. Nowadays, with the support of most DB Management Systems and CASE tool, it is easy to retrieve the DB Schema from a dataset or a database in relational form. A DB schema includes a set of tables or relations and links of foreign keys between tables. System analyst defines the correspondence of elements in DB schema and states in OLC. The general steps of our method are described in Figure 2. The first and second tasks are carried out manually; meanwhile the last phase can be done automatically. In the following we describe in the detail these tasks.

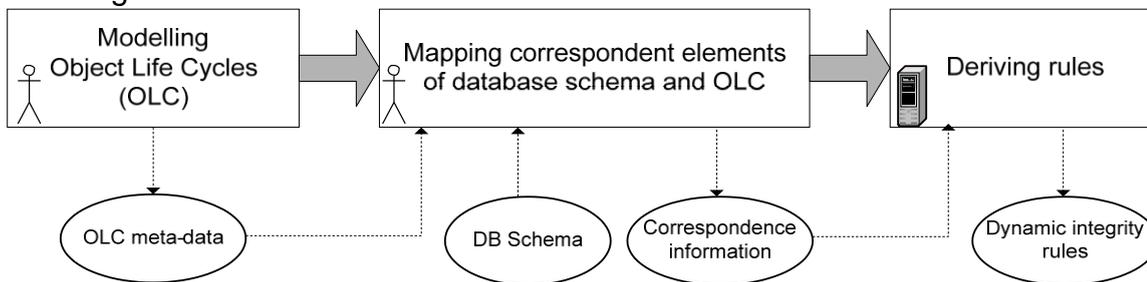

Figure. 2 Method for discovering dynamic IRs based on OLC

**OLC MODELLING**

The OLC model represents the dynamic aspect of our information system modelling concepts (PhamThi et al, 2005) (PhamThi, Helfert 2007a). An OLC is presented by a bipartite graph, which is called Node-Star Net. In this graph, a node corresponds to a state, and a star corresponds to a process.

A *state* is an object situation which satisfies some conditions. In it, objects execute methods or wait for participating into processes. A *process* corresponds to a system process, a business process, an administrative process, etc., or to a decision. A process owns its pre-condition and post-condition. A process is enabling if its pre-condition is satisfied, after a process is carried out, its post-condition is satisfied.

An OLC with the node-star structure is described as follows: OLC = <S, P, fi, fo, loop, back-inactive>

S: a set of states, P: a set of processes.

fi: an input function, fi : (S x P) →{0, 1}, fi(s, p) = 1 if s is an input state of the process p.

fo: an output function, fo : (T x S) →{0, 1}, fo(p, s) = 1 if s is an output state of the process p.

loop: (S x S) →{0, 1}, loop(sj, si) = 1 if sj is a looped state of si (si is a predecessor state of sj);

back-inactive: (S x S) →{0, 1}, back-inactive(sj, si) = 1 then when an object changes to the state sj, it leaves the state si, (si is a predecessor state of sj), otherwise object still keeps the state si, which results that objects resides in many states at the same time.

This model allows modellers to describe process flow thanks to the pre/post-conditions of processes and back-inactive function.

In Figure 3, we illustrate the OLC modelling of Order objects and DB schema within the e-commerce example. In this example, an *order* object leaves its current state when it changes to a new state. When a *Shipped* order object changes to *Closed* state, it leaves the *Shipped* state, therefore the *Return* process can not be carried out on this object (the pre-condition of the *Return* process is that there is an *order* object in *Shipped* state and a *return* object concerning this order), or vice-versa when a *Shipped* order object changes to *Returned* state. In other words this corresponds to an exclusive process control.

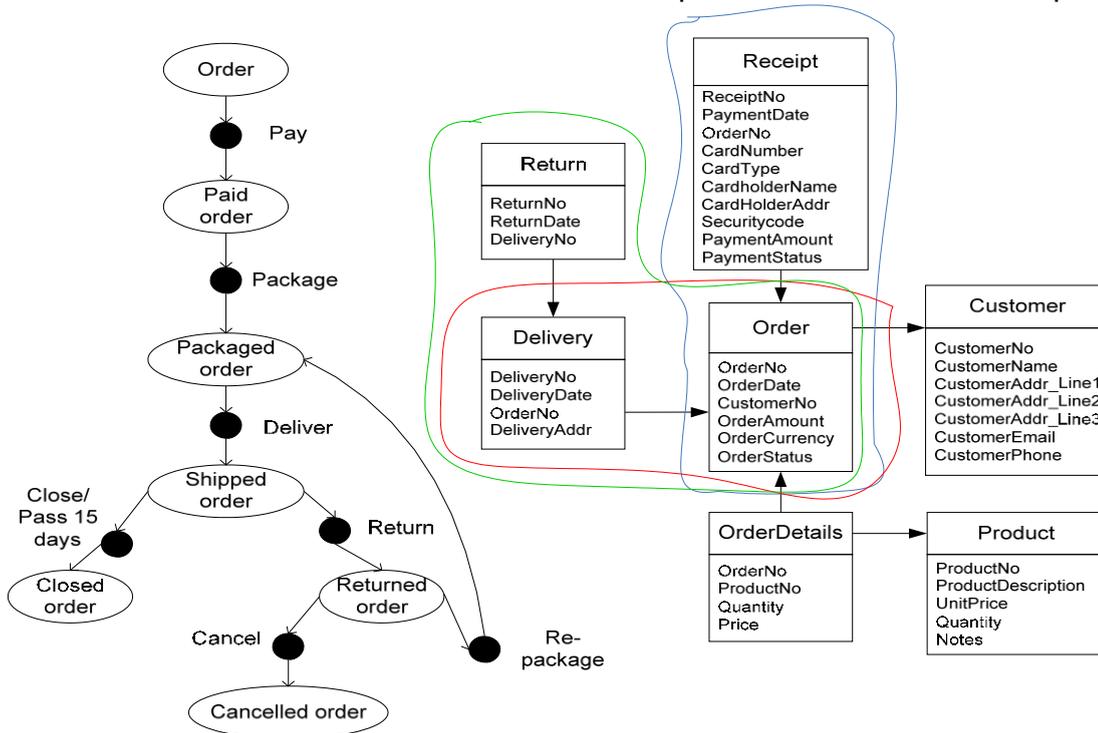

Figure 3. Object life cycle of Order and DB schema of E-commerce example

**Mapping correspondent elements of DB Schema and OLC states**

The OLC describes possible states of the objects in correspondent table(s) from the DB schema. These also represents in what order objects change their states. Determining the correspondences between OLC states and data elements in a DB schema helps to derive dynamic rules.

We identify following situations in a DB schema which correspond to states in OLC:
1. A state is defined as a table in a DB Schema
2. Explicit descriptions of object states with one or many attributes of a table in DB schema, in other words, the attribute values describe object states. For example, the *OrderStatus* attribute of *Order* table describes different order states such as *paid*, *packaged*, *shipped*, etc.
3. Implicit description of object states in DB schema which is usually based on links established between objects. For example, an order is changed to *Paid* state if there is a *receipt* object linked to this order and vice versa, or the order is changed to *Shipped* state if there is a *delivery* object linked to this order and vice versa, or the order is changed to *returned* state if there is a *return* object linked to a *delivery* object which is linked to that order object.
4. Combination of attribute values and links. This means an object state is determined by attribute values and links establishment from (to) this object to (from) other objects.

This correspondence information are specified and stored in a Correspondence repository.

**Deriving dynamic integrity constraints**

In a DB Schema, it is possible that the definition of various states of the same object may fall into various situations identified. For example with the same object, one state is defined by attribute values, but another state is defined with links between this object to other objects In any cases, dynamic IRs are derived with respect to states and the related order of state change specified in OLC. In this regard, the rule templates rest the same, but the detail is changed regarding how a state is defined. In the following, we present templates of derived rules based on this principals, and subsequently illustrate it with the current example:

*1. Value domain of status attributes must belong to the states specified in the OLC*

In the above example, suppose the *OrderStatus* attribute in *Order* table defines different states of an *order* object. Basing on its OLC, this rule, which is described in a similar form of predicate logic, is obtained:

- $\forall$order, order.OrderStatus $\in$("Paid", "Packaged", "Shipped", "Closed", "Returned", "Cancelled").

*2. The order of value change of status attributes must respect the order of state changes specified in the OLC*

In the above example, *OrderStatus* attribute can not change from, for example, "Paid" value to "Returned" value, or from "Shipped" value to "Paid" value:

- $\forall$order, if order.OrderStatus.New = "Shipped" then order.OrderStatus.Old= "Packaged", etc.

*3. The links establishments between objects must be consistent with the value of the status attributes, if applicable*

In the above example, basing on the correspondence specification in the previous section, we can obtain these rules:

- if order. OrderStatus="Paid" then $\exists$receipt, receipt.OrderNo= order.OrderNo, and
- if order.OrderStatus="Shipped" then $\exists$delivery, delivery.OrderNo= order.OrderNo, etc.

*4. The order of link establishments (and with or without attribute value) which represent different states must respect the order of state changes specified in the OLC.*

For example, the link between a *delivery* object to an *order* object must be established after the link between that *order* object to a certain *receipt* object. This rule can be described as follows:

- $\forall$order, if $\exists$delivery, delivery.Order = order.OrderNo then $\exists$receipt, receipt.OrderNo= order.OrderNo

*5. If there are exclusive states and they correspond to link establishments (and with or without attribute value) then these link establishments must be exclusive as well*

For example, in the above example "Closed" and "Cancelled" are exclusive states, if there is any links represent these states then they must be exclusive. Suppose there is a link establishment between a X table to the Order table which corresponds to "Closed" state, and there is a link establishment between a Y table to the Order table which corresponds to "Cancelled" state, then we can obtain this rule: If $\exists$x, x.OrderNo= order.OrderNo then $\forall$y, y.OrderNo $\neq$order.OrderNo.

**DISCUSSION**

Based on our method described above, we have developed a tool that includes the elements of the framework presented in Figure 1 for DQ analyzing with a dynamic integrity constraints generating function. We designed the OLC repository according to our Node-Star structure model, Correspondence repository for storing the correspondence definition between states in OLC and elements in DB schema. We also have developed algorithms for dynamic rules derivation based on defined rule templates. In this paper we do not present the structure of these databases and the algorithms.

With the support of the tool this systematic method is east to apply. Actually the business expert or system analysts solely need to specify the OLC of main business objects but not whole business process. Furthermore they need to map correspondent elements in DB schema to states in OLC before the deriving rules step can be automatically done. Therefore it is obvious that the effectiveness of this approach also depend on OLC modelling and state mapping. We have also discussed our approach with some DQ solutions providers; actually they have met challenges in specifying IRs because this has been manually done. Thus a systematic and semi-automatic approach sounds promising. Furthermore, this approach is efficient because it is based on OLC which is a part of business process and DB structure which is the root cause why such rules are needed. In case of evolving business, OLC may be changed and then correspondent rules may be changed as well; therefore this approach helps systematically managing rules.  The limitation of this approach is that business experts or system analyst must model OLC according to the Node-Star structure. This can be avoided if we develop OLC templates in different domain applications for references or reuse purpose.

**CONCLUSION AND FUTURE WORK**

DQ approaches based on BRs become increasingly promising. Current tools and BRs based approach have met challenges in discovering rules. However, most approaches are based on a manual process and significant domain expertise.  We have presented in this paper our rules-based approach for DQ analyzing. In this paper we particularly focus on describing a method for deriving dynamic rules based on the OLC concept. We also have developed a tool that implements our approach.  In the future we study constraints on different states of different OLC for automatic derivation. We also study importing existing business process and use them for this purpose.